\documentclass[11pt]{article}

\usepackage[top=2cm, bottom=2cm, left=2cm, right=2cm]{geometry}

\usepackage[noadjust]{cite}

\usepackage[cmex10]{amsmath}
\usepackage{amsbsy}

\usepackage{stackengine}
\usepackage{makecell}
\usepackage{boldline}
\setcellgapes{3pt}

\usepackage{algorithm}
\usepackage{algpseudocode}
\usepackage{varwidth}

\usepackage{tikz}
\usetikzlibrary{shapes, shadows, arrows}
\usepackage{url}

\usepackage{graphicx}
\usepackage{float}
\usepackage{stfloats}
\usepackage{wrapfig}

\usepackage{epstopdf}

\usepackage{amsthm,dcolumn,graphicx,multicol,fancyhdr}
\usepackage{latexsym}
\usepackage{graphicx}
\usepackage{ifthen}
\usepackage{rotating,amsfonts,color,amssymb,amsmath,amscd,array,enumerate,afterpage}

\usepackage{hyperref}
\usepackage{multirow}

\usepackage{enumitem}
\setlist{parsep=3pt,listparindent=\parindent}

\usepackage[caption=false,font=footnotesize]{subfig}

\usepackage{array}
\newcolumntype{L}[1]{>{\raggedright\let\newline\\\arraybackslash\hspace{0pt}}m{#1}}
\newcolumntype{C}[1]{>{\centering\let\newline\\\arraybackslash\hspace{0pt}}m{#1}}
\newcolumntype{R}[1]{>{\raggedleft\let\newline\\\arraybackslash\hspace{0pt}}m{#1}}

\usepackage{arydshln}

\theoremstyle{definition}

\newcommand{\babc}{\renewcommand{\labelenumi}{(\alph{enumi})}\begin{enumerate}}
\newcommand{\eabc}{\end{enumerate}}
\newcommand{\biii}{\renewcommand{\labelenumi}{(\roman{enumi})}\begin{enumerate}}
\newcommand{\eiii}{\end{enumerate}}

\newcommand{\beqn}{\begin{eqnarray*}}
\newcommand{\beq}{\begin{eqnarray}}
\newcommand{\eeqn}{\end{eqnarray*}}
\newcommand{\eeq}{\end{eqnarray}}

\newcommand{\ckboldon}[1]{#1}

\newcommand{\ckbold}[1]{%
 \ifthenelse{\isundefined{\ckboldon}}{#1}{ \textbf{#1} }
}

\begin{document}
\date{}
\title{Short-segment heart sound classification using an ensemble of deep convolutional neural networks}
\author{
Fuad Noman\footnote{School of Biomedical Engineering \& Health Sciences, Universiti Teknologi Malaysia, Malaysia (e-mail: mnfuad3@live.utm.my; hussain@fke.utm.my)},
Chee-Ming Ting\footnote{School of Biomedical Engineering \& Health Sciences, Universiti Teknologi Malaysia, Malaysia, and also the Statistics Program, King Abdullah University of Science and Technology, Thuwal, 23955-6900, Saudi Arabia(e-mail: cmting@utm.my)},
Sh-Hussain~Salleh\textsuperscript{*}, 
and Hernando Ombao \footnote{Statistics Program, King Abdullah University of Science and Technology, Saudi Arabia(e-mail: hernando.ombao@kaust.edu.sa)}} 

\maketitle

\begin{abstract}

This paper proposes a framework based on deep convolutional neural networks (CNNs) for automatic heart sound classification using short-segments of individual heart beats. We design a 1D-CNN that directly learns features from raw heart-sound signals, and a 2D-CNN that takes inputs of two-dimensional time-frequency feature maps based on Mel-frequency cepstral coefficients (MFCC). We further develop a time-frequency CNN ensemble (TF-ECNN) combining the 1D-CNN and 2D-CNN based on score-level fusion of the class probabilities. On the large PhysioNet CinC challenge 2016 database, the proposed CNN models outperformed traditional classifiers based on support vector machine and hidden Markov models with various hand-crafted time- and frequency-domain features. Best classification scores with 89.22\% accuracy and 89.94\% sensitivity were achieved by the ECNN, and 91.55\% specificity and 88.82\% modified accuracy by the 2D-CNN alone on the test set.

\end{abstract}

\vspace{5mm}

\vspace{-0.05in}

\section{Introduction}
\label{sec:intro}

Cardiac auscultation based on heart sound recordings or phonocardiogram (PCG) remains a primary screening tool for diverse heart pathologies. 
Various algorithms have been developed aiming at accurate automated classification of normal and abnormal PCGs \cite{Clifford2017}. However, the classification accuracy is still far from being reliable for diagnostics in clinical or non-clinical settings.
One major challenge is to extract robust and discriminative features from the raw PCG recordings typically corrupted by various noise sources. Different time-frequency and statistical features have been employed in automatic heart sound classification. Heart-rate variability is the most widely-used feature, which however can only be extracted from long recordings containing many cardiac cycles. Here we consider the challenges in obtaining high PCG classification accuracy for single individual cardiac cycles.
Recent developments in deep learning (DL) techniques have seen remarkable success in many practical classification tasks, sometimes surpassing human-level performance \cite{Dodge2017}. This is owing to its inherent mechanism integrating both feature extractor and classifier, which permits learning of complex data representations with hierarchical levels of semantic abstraction via its multiple stacked hidden layers and hence the robust and accurate pattern classification even based on raw data or primitive features. It offers substantial gain in accuracy over traditional linear and kernel methods with shallow architecture. One popular DL architecture is the convolutional neural network (CNN) which alternately stacks a convolutional layer to extract feature maps through sparse localized kernels with weight sharing, and a sub-sampling or pooling layer to acquire invariance to local translation. CNNs have achieved state-of-the-art performance in diverse challenging image recognition tasks \cite{krizhevsky2012,ronneberger2015,shelhamer2017}.

Applications of DL to cardiac signals are introduced very recently \cite{Zhang2017,Acharya2017,Potes2016}. 
CNNs have been used for normal/abnormal PCG classification using input features such as spectrogram and Mel-frequency cepstrum coefficients (MFCCs) in \cite{nilanon2016} on 5-second windowed segments, and MFCC heatmaps of 3-second segments in \cite{rubin2016}. Tschannen et al. \cite{tschannen2016} combined a wavelet-based deep CNN feature extractor with support vector machine (SVM) for heart-sound classification. Zhang et al. \cite{zhang2016segmental} proposed a segmental CNN model to detect cardiac abnormality with two different designs to adjust the configuration of convolutional layers filters. A DL architecture was implemented on field programmable gate array (FPGA) for real-time heart-sound classification using inputs based on gray sonogram images transformed from PCG segments \cite{dominguez2018}.

In this paper, we propose a deep CNN for classification of pathology in PCG of a single heart beat. We design a new architecture called time-frequency ensemble CNN (TF-ECNN) that combines a 1D-CNN and a 2D-CNN using respectively the time-domain raw PCG signals and MFCC time-frequency representations as inputs. Our method was evaluated on the PhysioNet computing in cardiology (CinC) 2016 challenge database \cite{Liu2016}, the largest heart sound database available so far. The aim is to classify the heart sound signal from a short segment (single cardiac cycle - heartbeat) into normal and abnormal classes.
We also investigated the performance of the proposed CNN model combined with different combination of input features, and compared with traditional classifiers, i.e., support vector machine (SVM), ensemble of decision trees and hidden Markov model (HMM). The hyperparameter tuning was carried out by using Bayesian optimization to find optimal values for model parameters \cite{Snoek2012} for all competing classifiers except HMM where the expectation-maximization algorithm was used to estimate the model parameters. 

\vspace{-0.05in}
\section{Methods}
\label{sec:methods}
\vspace{-0.05in}

In this section, we describe the main building blocks of heart sound classification algorithm consisting of preprocessing, segmentation, feature extraction and classification. For classification, we propose an ensemble of two deep CNNs that combines time-domain and frequency-domain input features, and consider three traditional approaches as baseline.

\vspace{-0.05in}
\subsection{Database}
\label{subsec:database}
\vspace{-0.05in}

We used heart sound recordings obtained from the PhysioNet CinC challenge 2016 database publicly available on PhysioNet website \cite{PhysioNet}. The dataset consists of 3153 recordings collected from healthy and pathological subjects. Recordings labeled as `unsure' by the cardiologists regarding the normal or abnormal categories were not used, leaving a total of 2872 recordings for training and evaluation in this work.

\vspace{-0.05in}
\subsection{Preprocessing}
\label{subsec:preprocess}
\vspace{-0.05in}

All the heart sound recordings were down-sampled to 1000 Hz and band-pass-filtered with Butterworth filter between 25 Hz and 400 Hz to eliminate the unwanted low-frequency artifacts (e.g., baseline drift) and high-frequency noise (e.g., background noise). The signals were then standardized by subtracting the mean and dividing by its standard deviation before feature extraction.

\vspace{-0.05in}
\subsection{Segmentation}
\label{subsec:segment}
\vspace{-0.05in}
The whole heart sound recordings were segmented into short intervals of single beat and then classified into normal and abnormal categories. In this work we used the heart sound annotations provided with the database for segmentation of each recording into heartbeats (from the beginning of atrial activity to end of ventricular activity). Note that other data-driven unsupervised algorithms such as the Viterbi alignment can also be used to perform such segmentation. A total of 81503 segments were extracted from the whole database which were then partitioned into subject-oriented train and test datasets with balanced number of samples as shown in Table.~\ref{Table:table1}.

\vspace{-0.05in}
\subsection{Baseline Classifiers}
\label{baselines}
\vspace{-0.05in}
 
We consider three baseline classifiers for comparison, namely, (1.) SVM with radial basis function kernel, (2.) ensemble of decision trees classifier and (2.) HMM.

\textit{SVM and desicion tree ensemble}. 
Following \cite{Potes2016}, a total of 58 features were extracted from each heartbeat for the SVM and ensemble of trees methods. These include 22 time-domain features (durations, skewness, kurtosis and sum of instantaneous amplitudes for each of the four heart sound states (S1, systole, S2 and diastole)) plus 36 frequency-domain features (median power spectrum for 9 frequency bands for each heart sound state). We further performed feature selection using the well-known neighborhood component analysis (NCA) \cite{Goldberger2005}, selecting a total of 28 features (16 time-domain and 12 frequency-domain).
We carried out 5-fold cross-validation to optimize and tune the hyperparameters of the SVM and the tree ensemble classifiers. A Bayes optimization approach is used to minimize the loss function and select the best set of hyperparameters that produce best classification results. We also applied class weights when computing the classification accuracy to accommodate possible misclassification of normal class, since the database is slightly imbalanced.

\begin{table}[!t]
\caption{Distribution of train and test set of the Physionet CinC challenge 2016 database.}
\vspace{0.1in}
\label{Table:table1}
\centering
\begin{tabular}{|l|c|c|c|c|}
\hline
\multirow{2}{*}{} & \multicolumn{2}{c|}{Train} & \multicolumn{2}{c|}{Test} \\ \cline{2-5} 
                  & normal      & abnormal     & normal     & abnormal     \\ \hline
Recordings        & 1150        & 284          & 1150       & 288          \\ \hline
Heartbeats          & 32574       & 8170         & 32582      & 8177         \\ \hline
\end{tabular}
\vspace{-0.1in}
\end{table}

\textit{HMM}.
Continuous HMMs with Gaussian mixture densities were used for modeling the temporal structure in PCG. We extracted a set of features as in \cite{Noman2018}.
A sequence of $12 \times 1$ short-time Mel-frequency cepstral coefficients (MFCCs) were computed over consecutive windowed frames for each heartbeat to obtain a two-dimensional $12 \times T$ time-frequency representation with $T$ the total number of feature vectors. 
A 4-state HMM with left-to-right topology was employed to model the time evolution of the four distinct heart sound components in a single heartbeat.
A mixture of 16 Gaussians was used as the observation model in each state. We found no practical improvement in classification accuracy for this data with larger number of Gaussian components. The HMMs were trained by using the Baum-Welch algorithm based on expectation-maximization to find the maximum likelihood estimates of the model parameters \cite{Rabiner1989}. The Viterbi algorithm was used for aligning the MFCC frames to each of the four cardiac states, and to compute the likelihood score of a test example which was then classified to the HMM with the highest likelihood.

\vspace{-0.1in}
\subsection{Proposed Ensemble CNN}
\label{subsec:cnn}
\vspace{-0.05in}
 
\begin{figure*}[th!]
  \centering
\begin{minipage}[b]{1.0\linewidth}
	\centerline{\includegraphics[width=18cm]{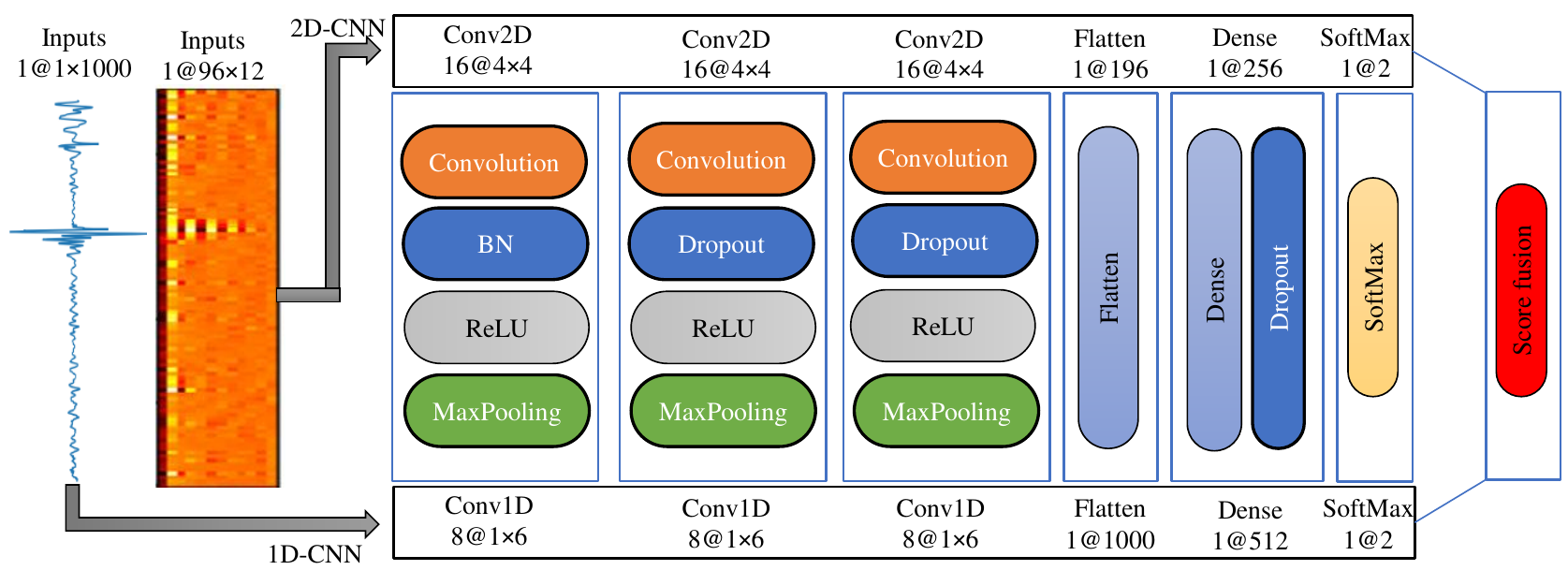}}
  \vspace{-.2cm}
  \caption{Architecture of the proposed TF-ECNN model combining a 1D-CNN and a 2D-CNN taking inputs of raw signals and time-frequency feature maps, respectively. BN: Batch-normalization layer. ReLU: rectified linear unit activation function.}
  \label{Fig:fig1}
\end{minipage}
\vspace{-0.2in}
\end{figure*}

Fig.~\ref{Fig:fig1} shows the architecture of the proposed time-frequency based ensemble deep CNN (TF-ECNN) model combining two distinct CNNs to capture the temporal structure in both the time-domain and frequency-domain. The first CNN (1D-CNN) accepts one-dimensional PCG time series data as input (i.e., the raw heartbeat signal). The second CNN (2D-CNN) uses the two-dimensional time-frequency feature maps of MFCCs and time-varying autoregressive (TV-AR) coefficients as input. For both the 1D-CNN and 2D-CNN, we used the same network architecture consisting of convolutional, activation, pooling and fully-connected (or dense) layers but with different sets of hyperparameters. 

\vspace{-0.1in}
\subsubsection{Feature Extraction}
\label{subsec:cnn-FE}
\vspace{-0.05in}

The 1D-CNN was designed to classify the raw heart sound from fixed-length segments. However, the heartbeat segments are usually with variable lengths. Therefore, two approaches were used to normalize the segments durations. First, an anti-aliasing linear interpolation method was performed to normalize the heartbeats into reference duration (i.e. 1000 samples). Second, the segments with durations higher that 1200 samples were ignored (1.2\% of total segments), then we zero-padded the rest of the segments to 1200 samples.

For 2D-CNN, we consider two approaches of feature extraction to obtain the two-dimensional time-frequency feature maps. First, similarly for the HMM classifier, we computed frames of short-time MFCC features on the duration-normalized PCG segments to produce feature maps of same size to represent each heart beat. Second, we computed autoregressive coefficients of 12-th order TVAR model commonly known as short-time linear predictive coefficients (LPCs) to construct alternative feature map for each segment.

\vspace{-0.05in}
\subsubsection{Network Architecture \& Training}
\label{subsec:cnn-models}
\vspace{-0.05in}

\begin{table}[!t]
\centering
\caption{Summary of 1D-CNN model configurations.}
\vspace{0.07in}
\label{Table: table2}
\resizebox{0.49\textwidth}{!}{
\begin{tabular}{lllcc}
\hline
Layer & Type        & Output shape                 & Kernel size & Strides \\
\hline
1     & Convolution & 1000 $\times$ 8 & 6           & 1       \\
2     & Batch-Norm  & 1000 $\times$ 8 & -           & -       \\
3     & MaxPooling  & 500 $\times$ 8  & 2           & 2       \\
4     & Convolution & 500 $\times$ 8  & 6           & 1       \\
5     & MaxPooling  & 250 $\times$ 8  & 2           & 2       \\
6     & Convolution & 250 $\times$ 8  & 6           & 1       \\
7     & MaxPooling  & 125 $\times$ 8  & 2           & 2       \\
8     & Flatten     & 1000            & -           & -       \\
9     & Dense       & 512             & -           & -       \\
10    & SoftMax     & 2               & -           & -      \\
\hline
\end{tabular}}
\vspace{-0.2in}
\end{table}

\begin{table}[!t]
\centering
\caption{Summary of 2D-CNN model configurations.}
\vspace{0.07in}
\label{Table: table3}
\resizebox{0.49\textwidth}{!}{
\begin{tabular}{lllcc}
\hline
Layer & Type        & Output shape                 & Kernel size & Strides \\
\hline
1     & Convolution & 96 $\times$ 12 $\times$ 16   & 4           & 1       \\
2     & Batch-Norm  & 96 $\times$ 12 $\times$ 16   & -           & -       \\
3     & MaxPooling  & 48 $\times$ 6 $\times$ 16    & 2           & 2       \\
4     & Convolution & 48 $\times$ 6 $\times$ 16    & 4           & 1       \\
5     & MaxPooling  & 24 $\times$ 3 $\times$ 16    & 2           & 2       \\
6     & Convolution & 24 $\times$ 3 $\times$ 16    & 4           & 1       \\
7     & MaxPooling  & 12 $\times$ 1 $\times$ 16    & 2           & 2       \\
8     & Flatten     & 192                          & -           & -       \\
9     & Dense       & 256                          & -           & -       \\
10    & SoftMax     & 2                            & -           & -      \\
\hline
\end{tabular}}
\vspace{-0.2in}
\end{table}


\begin{table*}[!ht]
\centering
\caption{Performance comparison of different classifiers on the test set. The numbers in parentheses correspond to the classifier performance before applying the weight cost for imbalanced classes.}
\vspace{0.05in}
\label{Table:table4}
\begin{tabular}{llcccc}
\hline\hline
Classifier           & Features     & Accuracy (\%) & Sensitivity (\%) & Specificity (\%) & MAcc (\%)     \\
\hline
SVM                  & Time \& Freq & 84.87 (85.09) & 85.82 (94.09)    & 81.09 (48.95)    & 83.46 (71.52) \\
\hline
Ensemble             & Time \& Freq & 86.20 (86.23) & \textbf{90.55} (94.25)    & 68.84 (54.26)    & 79.70 (74.26) \\
\hline
HMM                  & MFCC         & 87.07 (n/a)  & 85.97 (n/a)     & 91.45 (n/a)     & 88.71 (n/a) \\
\hline
\multirow{2}{*}{1D-CNN} & Raw (zero-pad) & 86.34 (85.63) & 87.80 (95.11)    & 80.32 (46.41)    & 84.06 (70.76) \\
												& Raw (norm-dur)    & 87.23 (87.52) & 87.57 (91.51)    & 85.84 (71.64)    & 86.71 (81.58) \\			
\multirow{2}{*}{2D-CNN} & TVAR         & 86.41 (86.91) & 88.85 (91.79)    & 76.69 (67.45)    & 82.77 (79.62) \\
												& MFCC         & 87.18 (89.30) & 86.08 (92.49)    & \textbf{91.55} (76.61)    & \textbf{88.82} (84.55) \\
ECNN & Raw (norm-dur) + MFCC & \textbf{89.22} (89.58) & 89.94 (93.07) & 86.35 (75.68) & 88.15 (84.37)\\
                     \hline\hline
\end{tabular}%
\vspace{-0.1in}
\end{table*}

Table \ref{Table: table2} and Table \ref{Table: table3} respectively summarize the architecture of the proposed 1D-CNN and 2D-CNN models individually. The experiments were carried out using TensorFlow platform \cite{Abadi2016} with Scikit-Optimize library which provides Bayesian optimization of the hyperparameters. We used the expectation-improvement method (with 100 iterations) to tune the CNN parameters, including learning rate, number of convolution layers, number of filters, kernel size, activation method, number of dense layers, number of nodes in dense layers, and dropout ratio of dense layers. We selected a fixed dropout ratio of 0.4 and 0.5 for the convolution layers of the 1D-CNN and 2D-CNN, respectively. All convolution layers used the zero-padding to preserve the input dimension. A batch-normalization layer was attached to the first convolutional layer to allow the model to learn different variations of the data which can give better robustness to noise typically present in real heart sound recordings. For other convolutional layers, we added dropout layer as regularization method to prevent model overfitting. 
 
Of notes, additional experiments showed that the use of zero-padding in input segments for the 1D-CNN did not perform as well as using the duration-normalized segments with the same CNN architecture. Thus, the dropout ratio of the convolution layers was set to 0.8. The Bayesian optimization procedure suggested a 2D-CNN architecture with similar number of convolution layers with the 1D-CNN but slightly different number of dense layers. Therefore, we manually tuned the 2D-CNN architecture to match that of 1D-CNN which performed comparably with the Bayesian-optimized model.
The learning rates set by the optimizer for the 1D-CNN and 2D-CNN were respectively 0.001031 and 0.000496 with batch size of 128. The Adam optimizer was used for weighs updating in the backpropagation training stage.

In the TF-ECNN, we combine both the 1D-CNN and 2D-CNN optimized above based on score-level fusion by summing over the outputs of softmax layers from two individual CNNs to produce fused class prediction probabilities.

\vspace{-0.1in}
\section{Experimental Results}
\label{sec:results}
\vspace{-0.1in}

We evaluate the classification performance of the 1D-CNN and 2D-CNN individually as well as the TF-ECNN, as measured by sensitivity, specificity and modified accuracy (MAcc). The MAcc is an average of the sensitivity and specificity scores. Table \ref{Table:table4} shows the results of different classifiers and feature sets on the local hidden-test set.


Numbers in parentheses indicate performance of trained models without using the weighed-cost to control imbalances among the classes. They show that all classifiers do not perform well with a significant tradeoff between the sensitivity and the specificity. This is due to the imbalanced classes and limited abnormal data which lead to a high misclassification of abnormal segments as clearly indicated by the specificity scores. After corrections by applying class weights to limit the misclassification of abnormal class, performance of all classifiers increases except the ensemble of tress (still with low sensitivity and MAcc of below 80\% but high sensitivity).

The proposed CNN models generally outperform the baseline classifiers considerably in most of the performance measures. In particular, the 2D-CNN with MFCCs achieved the best performance in specificity and MAcc, and the TF-ECNN gives the highest accuracy and the second highest in sensitivity. HMM follows, performing the best among the traditional classifiers, possibly due to capability of the Markov chain in modeling the temporal structure of the four heart-sound states which is neglected by the SVM and even the CNNs. The performance of the ensemble of trees is not well-balanced, scoring highest sensitivity but with the lowest sensitivity and MAcc.

It is interesting to note that the 1D-CNN using only raw-data as input shows a satisfactory performance compared to using computationally-expensive feature extraction methods (i.e., MFCC and TVAR) in the 2D-CNN. The 1D-CNN with duration-normalized raw PCG is only 2\% less than the best MAcc score obtained by the 2D-CNN with MFCC. This may suggest the advantages of the multiple hidden layers in CNNs that can learn hierarchical time-frequency features directly from the raw PCG signal. For the 2D-CNN with two-dimensional feature maps, the better time-frequency representation of the acoustic-based PCG signals using the MFCCs improves the classification performance over the TVAR. The ECNN combining both raw and MFCC features offer gains in sensitivity over the 2D-CNN using MFCC alone which however performs better in specificity, suggesting the advantage of ECNN in detecting the normal heart sounds whereas the 2D-CNN for the abnormal heart sounds.

\vspace{-0.15in}
\section{Conclusion}
\label{sec:conclusion}
\vspace{-0.1in}
We developed an ensemble of deep CNNs to classify normal and abnormal heart sounds based on short-segment recordings of individual heart beats with promising performance. The novel network architecture combines a 1D-CNN and a 2D-CNN designed respectively to learn multiple levels of representations from both the time-domain raw signals and time-frequency features.
Evaluation on large PhysioNet CinC challenge 2016 database demonstrates advantages of our proposed CNN models with considerable improvement in classification performance over strong start-of-the-art baseline classifiers and feature sets. This suggests potentials of deep learning approaches for accurate heart-sound classification. Future works will consider use of sequential DL models such as the recurrent neural networks (RNNs) or long short-term memory (LSTM) RNNs \cite{hochreiter1997} that could better capture the temporal dependency in the time-varying spectrum of PCG signals.


\bibliographystyle{IEEEbib}
\bibliography{refs}

\end{document}